# Title: Superlubricity of Graphene Nanoribbons on Gold Surfaces

**Authors:** Shigeki Kawai[1,2,*,†], Andrea Benassi[3,4,*,†], Enrico Gnecco[5,6], Hajo Söde[3], Rémy Pawlak[1], Xinliang Feng[7], Klaus Müllen[8], Daniele Passerone[3], Carlo A. Pignedoli[3], Pascal Ruffieux[3], Roman Fasel[3,9] and Ernst Meyer[1]

**Affiliations:**

[1]Department of Physics, University of Basel, Klingelbergstrasse 82, CH-4056 Basel, Switzerland.

[2]PRESTO, Japan Science and Technology Agency, 4-1-8 Honcho, Kawaguchi, Saitama 332-0012, Japan.

[3]nanotech@surfaces Laboratory, Empa, Swiss Federal Laboratories for Materials Science and Technology, Überlandstrasse 129, 8600 Dübendorf, Switzerland.

[4]Institute for Materials Science and Max Bergmann Center of Biomaterials, TU Dresden, 01062 Dresden, Germany.

[5]Instituto Madrileño de Estudios Avanzados en Nanociencia, IMDEA Nanociencia, 28049 Madrid, Spain.

[6]Otto Schott Institute of Materials Research, Friedrich Schiller University Jena, 07743, Jena, Germany.

[7]Department of Chemistry and Food Chemistry, Center for Advancing Electronics Dresden (CFAED), Technische Universität Dresden, 01062, Dresden, Germany.

[8]Max Planck Institute for Polymer Research, 55124 Mainz, Germany.

[9]Department of Chemistry and Biochemistry, University of Bern, Freiestrasse 3, 3012 Bern, Switzerland.

*These authors contributed equally to this work.

†Correspondence to: shigeki.kawai@unibas.ch and andrea.benassi@nano.tu-dresden.de

**Abstract**: The state of vanishing friction known as superlubricity has important applications for energy saving and increasing the lifetime of devices. Superlubricity detected with atomic force microscopy appears in examples like sliding large graphite flakes or gold nanoclusters across surfaces. However, the origin of the behavior is poorly understood due to the lack of a controllable nano-contact. We demonstrate graphene nanoribbons superlubricity when sliding on gold with a joint experimental and computational approach. The atomically well-defined contact allows us to trace the origin of superlubricity, unravelling the role played by edges, surface reconstruction and ribbon elasticity. Our results pave the way to the scale-up of superlubricity toward the realization of frictionless coatings.

**One Sentence Summary:** Graphene superlubricity is investigated with atomic scale precision on nanoribbons anchored and dragged by a sharp tip on a gold surface.

**Main Text:** Graphene offers unique properties as solid lubricant (*1*) and has a potential to be used as an ultra-thin coating material on surfaces, almost suppressing energy consumption in mechanical components. The key interpretation of such a so-called "superlubric" behavior is based on these facts (*2-5*): (i) The high lateral stiffness of graphene makes a commensurable contact with most solid surfaces hardly possible. (ii) Combined with the weak interaction with most materials, incommensurability leads to a state of ultralow friction when graphene slides over a different material. To substantiate this hypothesis, and establish a connection with the tribological properties observed on macro/meso-scales, it is highly desirable to measure the mechanical response of a graphene flake down to the nanometer level. In such measurements, one has to ensure that both contacting surfaces are atomically well-defined, their common interface is free from contaminants and the ultralow forces accompanying the sliding motion can be distinguished from the background noise. While a clean atomically flat surface as a substrate can reliably be obtained in ultra-high vacuum (UHV), atomically defined graphene systems as sliding objects are hardly prepared. Carbon nanotubes have exceptional superlubric properties up to a length scale of few cm (*6*), but their curvature makes them not easy to manipulate in a controlled way. Nevertheless, the problem can be overcome by employing graphene nanoribbons (GNRs), recently synthetized on a metal substrate by on-surface chemical reaction (*7*). Their structure is well-defined by the precursor molecule, as confirmed by high resolution scanning tunnelling microscopy (STM) and atomic force microscopy (AFM). For this reason, the GNRs are an appropriate candidate for our goal. Apart from that, GNRs are also very promising in a series of applications (e.g. nano-electromechanical systems (*8*), nanofillers (*9*), transistors (*10*), and other electronic and spintronics devices (*11*)) where assessing their mechanical stability is pivotal.

We investigate the frictional, adhesive, and elastic properties of GNRs by lateral manipulation on an Au(111) substrate by using dynamic AFM in UHV at low temperature (4.8 K). The end of selected GNRs was accurately anchored to the probing tip and dragged back and forth in a controlled way while the friction force was recorded. An accompanying computational "experiment" allows us to relate the origin of superlubricity so measured to the molecular dynamics occurring at the interface.

Our measurements originate from the accidental manipulation of GNRs aligned along the [-1,0,1] direction of the Au(111) substrate, when the GNRs were imaged by STM using a gold tip. The GNRs were always displaced along their longitudinal axis even with a relatively large separation, indicating high diffusivity (fig. S1-S4). To measure the static friction ($F_{\text{stat}}$) we switched to AFM using the same tip. After imaging a sample area covered by GNRs (Fig. 1A) we acquired a two-dimensional (2D) frequency shift map while the tip was scanned laterally along the GNR (*X* direction) at different constant Z distances (Fig. 1B and fig. S5). Following the method of Ternes *et al.* (*12*), we reduced the tip-GNR distance stepwise during scanning until we observed an abrupt decrease of $\Delta f$ at the distance defined as Z=0 (Fig. 1C). We found the GNR displaced by a distance *d*=2.2 nm (Fig. 1A). Interestingly, we observed that the $\Delta f(X)$ profile is repeated after the same distance. Langewisch *et al.* reported a similar observation in their manipulation experiments on PTCDA molecules (*13*). We found that *d* varied with both the GNR length and the adsorption site. Furthermore, jumps with smaller *d* values were rarely observed, and the GNRs were never dragged continuously, meaning that the junction formed between tip and GNRs is rather weak. To quantify $F_{\text{stat}}$, we first estimated the energy landscape experienced by the tip by integrating two times the $\Delta f(Z)$ sections extracted from Fig. 1B. Then, we differentiated the 2D potential map along the *X* direction and multiplied the result by the factor $-2k_c/f = -0.15$ Nm$^{-1}$Hz$^{-1}$, where $k_c$=1800 N/m is the spring constant and $f$=24.7 kHz is the resonance frequency of the free tuning fork. The manipulation occurred when $F_{stat} \approx -105$ pN (Fig. 1D). Note that $F_{stat}$ is exceptionally low, considering that the linear size of the GNR is well above those of single atoms and conventional molecules which are typically manipulated by AFM (*12,14,15*). This result is a strong confirmation of the superlubric properties of graphene on the nanoscale, as observed in previous friction measurements on graphene flakes with undefined size (*16-18*). Another signature of superlubricity is the decrease of the friction force per unit contact area with increasing size of the contact (*19-21*). To this aim, in our quasi-1D system, we have repeated the measurements on GNRs of different length (Fig. 1E). In spite of the great dispersion in the measured data (due to the surface reconstruction, see below) the force per unit length is indeed found to decrease with increasing GNR length.

We estimated the diffusion barrier for the GNR assuming a simple sinusoidal interaction as $\Delta E \approx F_{stat}\, a/\pi \approx 40$ meV, where $a$=0.41 nm is the lattice constant of the Au(111) substrate (*15*). This remarkably low value means that single, isolated GNR would diffuse spontaneously at room temperature, making measurements challenging. We have also observed a rotation of short (2 nm) GNRs, although we always scanned the tip exactly along the GNR axis (fig. S6). This behavior is predicted theoretically for graphene flakes dragged on graphite (*22*). We even observed a vertical motion of the shortest GNRs (1 nm) before the start of lateral manipulation (fig. S7). We could not perform reliable static friction force measurements on GNRs longer than 22 nm because other GNRs were often found in the proximity, and the measured forces were considerably affected by the interaction with those neighbors. Yet, superlubricity allowed us to manipulate GNRs up to 55 nm long (fig. S8 and S9).

Although our measurements allow a precise estimation of the static friction force, they do not provide any details on the complex dynamics of the sliding motion of the GNRs. To gain more insight, we applied the procedure introduced by some of us for polymer chains (*23*) and succeeded in attaching a short edge of a GNR to an Au coated tip. We then oscillated the GNRs along the [−1,0,1] direction of the Au(111) surface with the tip kept at a constant distance (*Z*) from the substrate. We repeated the measurements several times at increasing values of *Z* (Fig. 2A). The corresponding variations of $\Delta f$ are shown in Fig. 2C-F for increasing values of *Z*. We found the frequency shift to oscillate with a periodicity of 0.28 nm, except when the GNR was driven backwards with a tip-surface distance *Z*=5 nm. The amplitude of the $\Delta f$ oscillations is not constant along *X*, but modulated on distances of few nm, where it varies by a factor of 2. Quite interestingly, we observed curves with roughly half periodicity at a small scanning distance of *Z*=1-2 nm (fig. S10 and S11). We also imaged the sample at the end of the process to ensure we manipulated only the target ribbon (Fig. 2B and fig. S12).

The simulated variations of the normal force $F_Z$ and lateral force $F_X$ as the tip drives the GNR parallel to the unreconstructed Au(111) surface at a low separation *Z*=2 nm are shown in Fig. 3 and fig. S13. Two characteristic lengths of 0.06 nm and 0.11 nm correspond to the lateral shift between three stable configurations (Fig. 3C). We estimated $\Delta f$ recorded in the AFM measurements by multiplying the force derivative $[F_Z(X, Z+\Delta Z)-F_Z(X, Z)]/\Delta Z$ by the conversion factor 0.15 N/m ($\Delta Z$=0.05 nm). The obtained profile (Fig. 3B) maintains the same periodicities of the manipulation curves, allowing comparison between simulations and experiments, although the relative heights of the peaks are different.

The regular profile (Fig. 3B) is considerably modified by the herringbone reconstruction, which deforms the top Au(111) layer and makes it slightly wavy (0.02 nm corrugation). The commensurability degree between GNR and substrate is modulated correspondingly, and the same modulation appears in the friction (or frequency shift) profiles. We have studied via our simulations the effect of the reconstruction starting at three different locations on the surface (Fig. 3D-G). The friction force is reduced if the whole GNR lies on the FCC or HCP regions (red arrows), is still small when the GNR crosses the boundary between FCC and HCP regions (green arrow), but it increases and reaches a maximum value when the free edge of the GNR or the point of detachment from the substrate sit over the boundary between FCC and HCP regions (blue arrows). The GNR short edge binds more strongly to the substrate than the inner atoms, and this effect is more pronounced in the boundary regions between HCP and FCC, where the substrate structure gets more commensurate to the GNR one. We further support this behavior by additional measurements on different GNRs (fig. S10) and theoretical studies on the pinning role of the nanostructure edges (*24*). The role of the herringbone reconstruction is confirmed by a similar experiment that we attempted on the unreconstructed Ag(111) surface. In this case the GNRs merged and formed a moiré pattern with the substrate (fig. S14). Manipulation with force values similar to those used on Au(111) was not possible in this case.

Configuration 2 of the GNR becomes unstable as the transition 1→2 and 2→3 are suppressed. This leads to the half periodicity we observed sometimes while scanning at very close separations. Just before a "slip" occurs the GNR becomes almost insensitive to the substrate, except for its short edge, still attached to the substrate (fig. S15). Because a C atom at this edge lies in a potential well $U_0$ a few meV deep and it is essentially driven by the spring $k=1.5$ N/m connecting the GNR to the tip apex (C-C bonds have an estimated stiffness of few hundreds N/m (*25*)) we can apply a well-known result of the Prandtl-Tomlinson model for atomic-scale friction and estimate the characteristic parameter $\eta=4\pi^2\ U_0/(ka^2)$ (*26*). The resulting value of $\eta$ is well below 1 and indicates a continuous transition between the two equilibrium states. Of course, this is strictly valid at the particular instant that we have considered. When the GNR is pinned in the configuration 1 or 3 and is pulled by the spring at the same time, all C atoms in contact with the substrate oppose a certain resistance, but the overall value of the friction force remains very small (a few hundreds of pN). Thus, our MD simulations are fully consistent with the commonly accepted interpretation for superlubricity of graphene. Due to its exceptional lateral stiffness this material is not prone to stretch and adapt to the substrate lattice while sliding. Combined with the weak interaction between graphene and substrate, the resulting incommensurability leads to the almost "frictionless" sliding of the GNR.

We plotted Δ*f(X)* curves at increasing separations *Z* from the surface and also reversed the direction of motion (Fig. 4). Configuration 1 becomes unstable when *Z*>2 nm, and only the 3→3 transitions remain (corresponding to the more frequently measured periodicity of 0.28 nm). Finally we notice that the forward and backward scan traces can be either in phase or in anti-phase. MD simulations allow to attribute this effect to the different bending of the suspended portion of the GNR in the two directions (Fig. 4). In the substrate regions with large friction force the bending of this portion can be much larger when scanning backwards, thus leading to a delay in the slip events (fig. S16). If *Z*=5 nm, i.e. close to the complete detachment, the agreement between model and experimental results becomes weak, which is presumably due to the fact that the H atoms passivating the GNR edges, neglected in the MD simulations, start to play an important role.

The pinning and releasing processes occurring in a sliding contact as shown here are pivotal in the development of friction between two solid surfaces in reciprocal sliding (*27*). The

GNR-Au(111) contact is almost superlubric, having static and kinetic friction force values in the range of 100 pN. The detailed dynamics of the sliding motion is nevertheless influenced by local surface properties, such as the variable degree of commensurability caused by the surface reconstruction. These details are clearly observable when the tip-surface separation Z is small, but tend to disappear as Z increases and the bending (elastic) properties of the suspended piece of GNR starts to play an important role. Our findings will help understanding and improving AFM based nano-manipulation techniques, and motivate the design of novel nano-functionalized interfaces for friction control.

**Acknowledgments:** This work was supported in part by the Japan Science and Technology Agency (JST) "Precursory Research for Embryonic Science and Technology (PRESTO)" within the project "Molecular technology and creation of new function", by the NCCR "Nanoscale Science" programme, by the grant CRSII2 136287/1 of the Swiss National Science Foundation, by the Swiss Nanoscience Institute and by COST Action MP1303, by EC under the Graphene Flagship (no. CNECT-ICT-604391), by Office of Naval Research BRC Program and by Comunidad de Madrid within the MAD2D-CM (S2013/MIT-3007) project. Part of the computational resources have been provided by the PRACE project 2012071262. All data in the main text and the supplementary materials are available online at www.sciencemag.org.

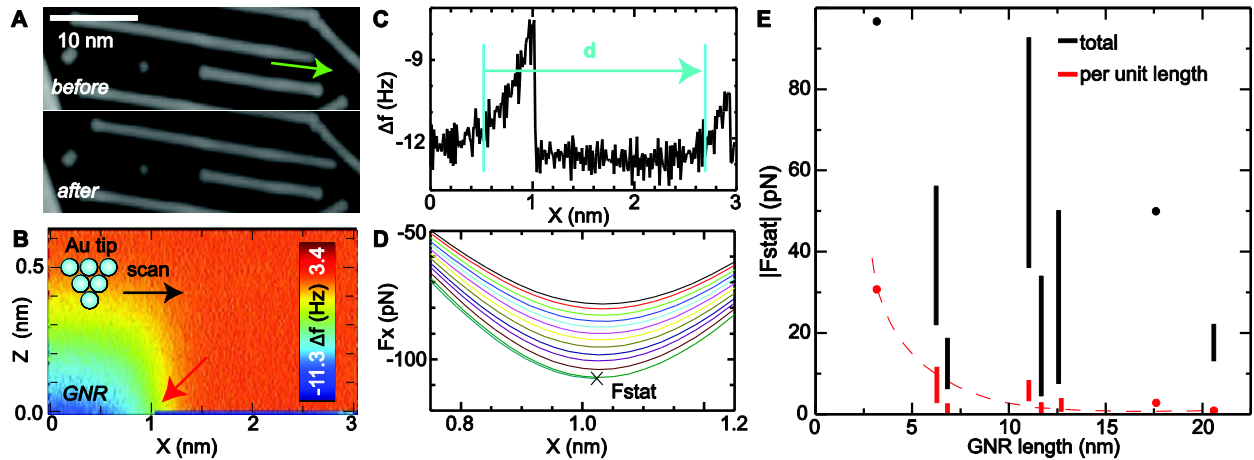

**Fig. 1**. Static friction force measurement. (**A**) STM topographies of GNRs on Au(111) before and after a tip-induced lateral manipulation (the green arrow indicates the sliding direction). (**B**) Two dimensional $\Delta f$ map along the longitudinal axis of the manipulated GNR. (**C**) Distance dependence of $\Delta f$ before, during and after the GNR displacement. (**D**) Calculated lateral force. The cross symbol corresponding to the red arrow in (B) shows the position at which the GNR starts moving and the corresponding value of the static friction force $F_{stat}$. (**E**) Absolute value of the static friction force $F_{stat}$ as a function of the GNR length (black) and $F_{stat}$ per unit length (red). Dots correspond to single measurements whereas bars connect the largest and the smallest values measured while manipulating the same ribbon on different surface regions. Measurement parameters: tunneling current $I=2$ pA and bias voltage $V=-200$ mV for (A) and oscillation amplitude $A=34$ pm for (B-D).

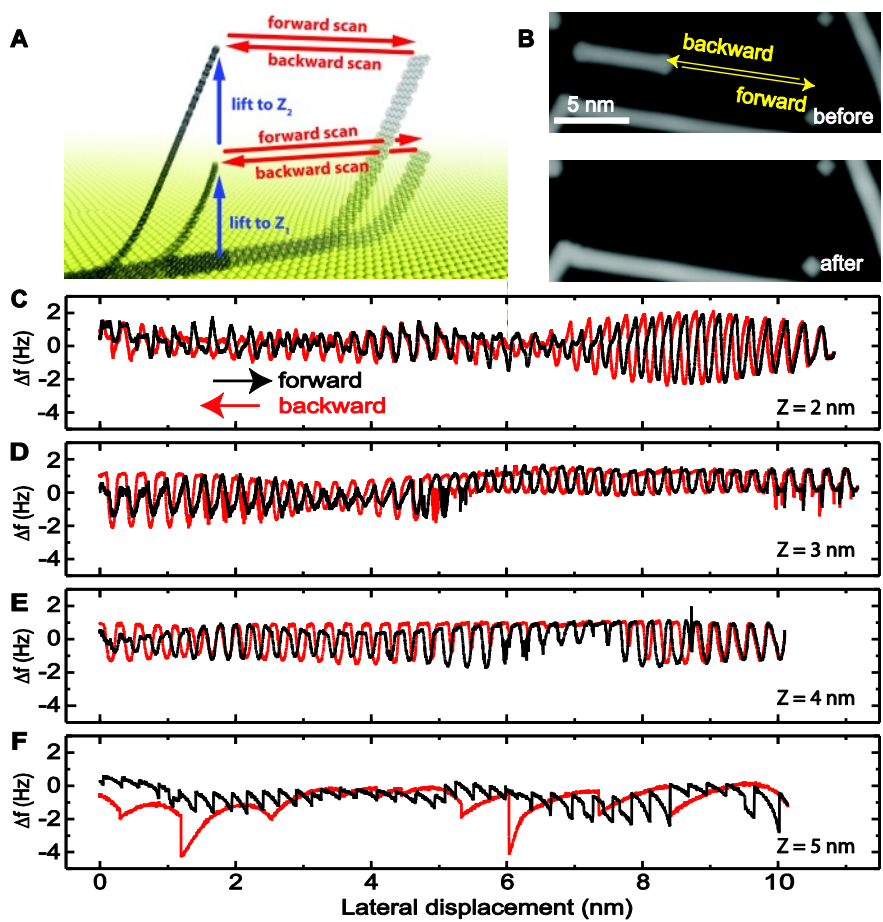

**Fig. 2.** Frequency shift versus pulling height. (**A**) Schematic drawing of the lateral manipulation procedure and (**B**) STM topographies before and after a GNR has been displaced on the Au(111) surface in the direction of the yellow arrows. The length of the GNR is 6.28 nm corresponding to 7 connected monomers. (**C-F**) Frequency shifts accompanying the lateral motion at different heights ($Z$=2,3,4,5 nm). Oscillation amplitude $A$=38 pm.

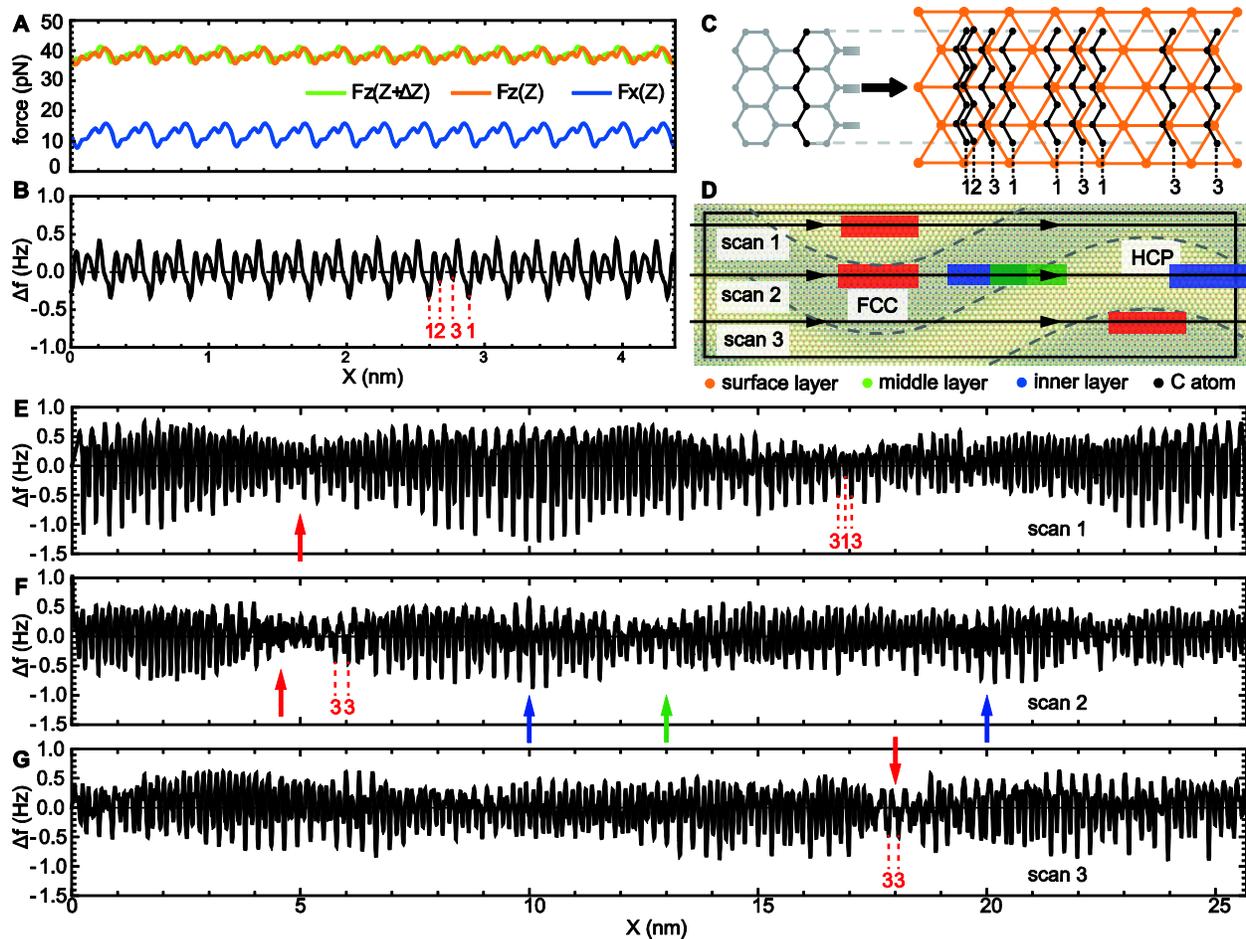

**Fig. 3.** Simulated sliding behavior. (**A**) Lateral force $F_X(X,Z)$ (blue curve) and normal force $F_Z(X,Z)$ (orange curve) while pulling the 6.28 nm long GNR along its longitudinal axis at a distance $Z=2$ nm from an unreconstructed Au(111) surface. The force $F_Z$ has been also calculated at $Z=2.05$ nm (green curve), which allows to estimate the frequency shift variation $\Delta f(X)$ of panel (**B**). (**C**) Sketch of a generic row of C atoms in the GNR (black) showing that the atoms sit most of the time in three non-equivalent configurations marked as 1, 2 and 3 originating a periodicity of ~0.06 nm for short jumps 1→2 or ~0.11 nm for long jumps 2→3, 3→1 (Au atoms are orange colored). (**D**) Tip trajectories on reconstructed Au(111) for the scan of panels and GNR configurations corresponding to minimum and maximum friction (the simulation cell size is 25.7×7.0 nm$^2$). Dashed lines represent the boundaries between HCP and FCC regions, rectangles represent the attached portion of GNR during the scan (with red, green and blue colors corresponding to increasing lateral force). (**E-G**) Frequency shift $\Delta f(X)$ along the scan lines in panel (D). The corresponding lateral force profiles are shown in fig. S13.

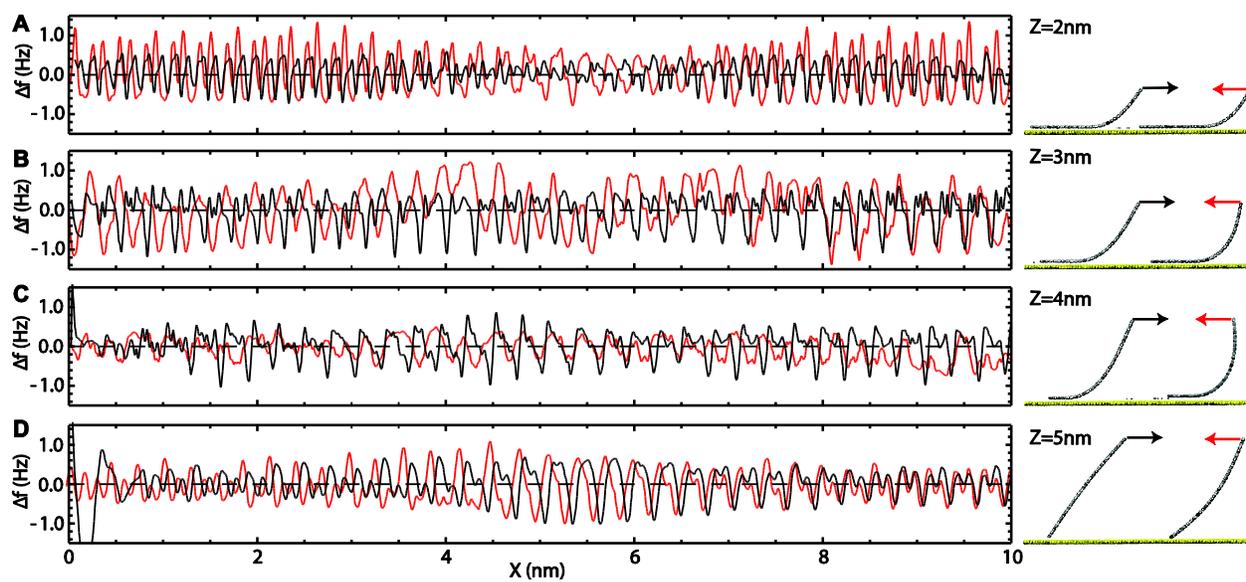

**Fig. 4.** Simulated frequency shift at different pulling height. (**A-D**) Frequency shifts for forward and backward scans at $Z$=2,3,4 and 5 nm. Note that the half periodicity disappears if $Z$>2 nm. The lateral insets show the different bending of the detached portion of GNR in the forward and backward scans. The corresponding friction force loops are shown in fig. S16.

# Supplementary Materials for

## Superlubricity of Graphene Nanoribbons on Gold Surfaces

Shigeki Kawai, Andrea Benassi, Enrico Gnecco, Hajo Söde, Rémy Pawlak, Xinliang Feng, Klaus Müllen, Daniele Passerone, Carlo A. Pignedoli, Pascal Ruffieux, Roman Fasel and Ernst Meyer

correspondence to:  shigeki.kawai@unibas.ch and andrea.benassi@nano.tu-dresden.de

**Materials and Methods**

Experimental

All experiments were performed with Omicron STM/AFM with a qPlus configuration (*28*), operating at 4.8 K in UHV. Clean Au(111) and Ag(111) surfaces were in-situ prepared by repeated cycles of standard sputtering and annealing. The tungsten tip of a tuning fork sensor was ex-situ sharpened by focused ion beam milling technique and was then in-situ covered with Au or Ag atoms by contacting to the sample surface. The resonance frequency of the self-oscillating qPlus sensor was detected by digital PLL circuits (Nanonis: OC4 and Zurich Instruments: HF2LI and PLL). 10,10'-dibromo-9,9'-dianthryl were deposited on the substrate from a crucible of Knudsen cell, resistively heated at 135 °C. Subsequently, the samples were annealed at 200 °C and 400 °C to synthesize graphene nanoribbons on Au(111) and Ag(111). The STM topographic images were taken in constant current mode. AFM imaging was performed in constant height mode, with the tip apex eventually decorated by a CO molecule.

Modeling

Molecular dynamics (MD) simulations have been conceived as an extension of the Frenkel-Kontorova model (*29*) to a 3D system including the proper GNR elasticity properties and the correct geometries of the Au surface and GNR lattice. The hydrogen atoms terminating the GNR edges have been neglected. In such a minimalistic model, only the very last atomic layer of the gold surface has been simulated, keeping the atomic coordinate fixed in time, therefore just providing a surface potential with the proper symmetry. For the unreconstructed Au(111) surface we used the experimental bulk lattice spacing 4.08 Å. The reconstructed surface, displaying FCC and HCP regions separated by elbow boundaries, was obtained following the procedure suggested by Narasimhan *et al.* (*30*). First, the DFT-derived coordinates for the 22×√3 reconstruction of Au(111) (*31*) was used as a starting building block. The slab is made up by 5 (111) layers fully optimized with the PBE/vdwDF unctional, (*32*) for a total of 266 atoms. The slab is then rotated around the axis normal to the surface by 60 degrees, and by attaching its mirrored image and appropriately duplicating in the *X* and *Y* direction, and removing too close atoms, an initial unit cell with a "v-shaped" profile, similar as the one proposed by Narasimhan *et al.* on the basis of STM experiments is obtained. MD using the embedded atom potential by Foiles *et al.* (*33*) is then applied to the slab in an orthorombic cell with periodic boundary conditions, for a total of 15061 atoms. Annealing runs at 500 and 250 K were performed for 4 and 1 ps, respectively, using a Nosé thermostat, followed by a quenching leading the system toward a structural minimum. The obtained configuration showed a typical dislocation pattern close to the elbow similar to the one observed in STM measurements.

The MD simulations have been performed using the LAMMPS code (*34*). The system is evolved through a Velocity-Verlet algorithm with time step of 1 fs. The interaction between carbon atoms has been modeled through both Tersoff and 2nd generation REBO potentials (*35,36*) finding no significant differences in the GNR equilibrium structure and mechanical response. Note that these potentials have been extensively used in previous studies to describe the mechanical properties of carbon based nanostructures (*37-39*). The gold-carbon interaction is represented by a Lennard-Jones potential. The simplifications introduced result in only two free independent parameters to fit the experimental data, namely the potential depth $\varepsilon$ and the characteristic length $\sigma$. A similar simple strategy has been employed to study other non-equilibrium phenomena such as friction and diffusion of gold clusters on graphite (*40,41*). The length $\sigma$ sets the GNR adsorption distance, and the choice $\sigma$=2.74 Å allows us to reproduce the value of 3.2 Å obtained from density functional theory calculations with van der Waals corrections (*42*). $\varepsilon$=2.5 meV has been chosen in order to reproduce the amplitude of the measured frequency shift oscillations and its dependence on the lifting height *Z*. Note that our choice of $\varepsilon$ is consistent with the upper bound value of 13 meV estimated experimentally (*43*). The initial rest configuration of the GNRs has been prepared placing them randomly on the surface and performing a slow annealing, decreasing linearly the temperature by a Langevin thermostat. With an annealing rate greater than $10^5$ K/fs the results of the simulations are found to be independent from the annealing protocol.

The dragging simulations have been performed driving the central carbon atom of one short edge, of coordinate *r*(*t*), at constant velocity $v_0$ through a spring *k* accounting for the combined torsion of the cantilever beam and lateral deformation of the tip apex. The normal and lateral (friction) forces have been calculated projecting the vector $F_{driving}(t) = k[r(t)-v_0 t\hat{x}]$ along the $\hat{z}$ and $\hat{x}$ directions respectively. A viscous damping term $m\gamma v$ has been added to each carbon atom (*m* atom mass and *v* atom velocity) to dispose off the energy injected into the GNR by the external driving force. In the experiments this energy flows away from the GNR into the gold substrate through the excitation of phonons and electronic degrees of freedom. *k*=1.5 N/m has been chosen to properly fit the experimental frequency shift magnitude. Note that it is very close to the value estimated in our previous work on polyfluorene chains (*23*) with a completely independent fitting procedure. A well-known issue in the simulations of nanoscale friction it that the choice of $v_0$ is limited by the need of sampling both the fast phonon/molecular vibration dynamics and the slower sliding in a reasonable simulation time (*44*). Taking into account the possible occurrence of stick-slip instabilities, a satisfactory (and quantitative) description of the tribological properties can be achieved with a choice of viscous damping $\gamma$ (10 ps$^{-1}$) and driving velocity $v_0$ (0.25 m/s) that decouples the fast atomic motion from the slower slider dynamics, even if the resulting $\gamma$ and $v_0$ values are orders of magnitude far from the experimental ones.

**Supplementary Text**

Accidental manipulation and diffusion barrier estimation

In this section we illustrate more deeply the GNRs accidental manipulation during STM imaging. Analyzing the GNRs behavior in different circumstances proved to be useful to draw some general qualitative conclusion about their interaction with the Au(111) surface.

We characterized the GNRs arrangement on Au(111) with a gold STM tip. GNRs align preferentially along the [-1,0,1] direction of the substrate (fig. S1A). Movement of some GNRs along their longitudinal axis during accidental manipulation at large tip-sample separation was the first indication of high diffusivity. We precisely determined the chemical structure of the GNRs using a CO functionalized AFM tip (*45*). We see carbon rings with long edges in an armchair type structure (fig. S1B). The low reactivity and smaller attractive forces of the CO tip compare to the metal tip prevented accidental manipulation during AFM imaging.

Several consecutive images of the same GNR are presented (fig. S2). The scattering noise visible at the termini of the GNR indicates the accidental manipulation of the GNR over several Au lattice sites caused by the probing tip (fig. S2A-C). Since the surface was imaged with the same parameters and the same tip conditions, in fig. S2D and E, where no sliding occurs, the GNR must be in a more energetically stable configuration. This accidental manipulation experiment thus reveals a site-dependent diffusion barrier.

The main reason for that is the Au reconstruction modifying the structure of the very last atomic plane and introducing a height modulation of the surface. Still fig. S3 shows how, due to the major pinning role played by the edges, also the GNR length might contribute significantly in determining the local diffusion barrier. In the figure two GNRs of slightly different length are found to lie parallel on the same region of the elbow reconstruction. The shorter one (ii) is entirely contained within the fcc region and is easily manipulated whereas the longer one (i) has the short edges sitting on the bridging region between hcp and fcc and it does not slide during imaging. We are thus led to the conclusion that the GNR edges bind much stronger to the bridging regions of the reconstructed surface, so the overall GNR length and its position relative to the reconstruction strongly affect its energetic and frictional properties.

The accidental slips performed by a GNR during imaging can be observed in more details in fig. S4. A first slip event (marked with a arrow) occurs on a length of about ~1.8 nm (a arrow). After a few scan lines a second shorter slip backward is recorded (b arrow) of about ~1.0 nm. Again the irregularity of these slips is due to the extremely complex energy landscape depending on position and length of the GNRs. To conclude this section we notice that the lateral manipulation of very short GNRs was not possible, since the GNRs jumps onto the tip apex if the last one is put very close (fig. S7). This means that the diffusion barrier is larger than the desorption one, which can be understood considering the high degree of commensurability with the substrate lattice achievable in this case.

Static measurements

In what follows we detail the static friction force measurements presented in Fig. 1 of the main text, discussing the difficulties arising when dealing with extremely long or very short ribbons. We start with fig. S8 showing a set of subsequent, controlled tip-induced slips through which an estimation of the static friction force is possible following the procedure described in the main text. Notice that the fast scanning direction was aligned parallel to the longitudinal axis of the GNRs. Two GNRs are manipulated: (i) with a length of 12.6 nm (15 precursor molecules) and (ii) with a length of 26.9 nm (32 precursor molecules). The images have been taken after every manipulation event. The slip direction and the measured static friction force of every event are directly shown in fig. S8. As discussed in the main text, even if the static friction force shows a large data dispersion independently of the length of the manipulated GNR, one can roughly say that the shorter GNR requires about half of the force necessary to induce a slip than the longer one. The shorter GNR is also found to undergo accidental manipulation during imaging (inset of fig. S8G).

Estimating the static friction force for shorter GNRs turned out to be a quite difficult task due to the dominant role played by the edges. For such GNRs the sliding motion is almost always accompanied by a rotation as highlighted in fig. S6, thus the estimated lateral force is no longer a measure of the static friction force only. Roto-translations occur systematically independently of the absolute position of the GNR with respect to the reconstruction and is found to disappear for GNRs with length ≥3 nm (i.e. composed at least by 3 monomers).

With very long GNRs the problem is different, since it is extremely difficult to find them entirely lying on a clean and free portion of surface without touching nearby GNRs or unreacted precursor molecules. A typical situation is shown in fig. S9 where a 55.5 nm long GNR (composed of about 66 precursor molecules) is manipulated in the direction indicated by the white arrow. Its edges are in contact with surrounding GNRs that, although not covalently bonded, provide an extra source of pinning spuriously increasing the measured lateral force.

Dynamical measurements

Here we provide extra details about the lateral dragging of the GNRs and the atomic scale features of their sliding motion. Fig. S10A represents a typical experiment: after the junction between the tip and the terminus of the GNR is established, the tip is retracted by 1 nm in the $Z$ direction, and subsequently laterally moved in the direction indicated by the yellow arrow. The frequency shift shown in fig. S10B is recorded during the scan and a final control image is taken after the manipulation to ensure that only the targeted GNR has been manipulated. As discussed in the main text the frequency shift shows certain characteristic periodicities, related to the atomistic matching of the contacting materials, and a larger scale modulation attributed to the Au reconstruction. In this manipulation the GNR is dragged until it hits another ribbon standing along its way, which results in the sudden jumps visible around 14 nm (fig. S10B).

Furthermore as discussed in the paper and supported by the simulations, when the GNR is lifted up from the surface at $Z<2$ nm, it is quite frequent to encounter wiggles with half periodicity with respect to the Au one. This correspond to the occurrence of 3→1→3 jumps. Evidences of this double periodicity are visible in Fig. 2C of the main text where the scan has been performed with $Z=2$ nm. Here we present a more pronounced evidence obtained scanning with $Z=1$ nm. Fig. S11 represents a zoom in the frequency shift signal reported in fig. S10. The expected standard Au periodicity of 0.28 nm is marked with dashed vertical lines, between them a second regular dimple is always present representing the intermediate configuration *1*.

The last point to be illustrated, to supplement the results of the main paper, is the GNR detachment occurring for large tip-surface separations $Z$. After the lateral manipulation reported in Fig. 2E of the main text, the tip has been retracted further up to 5.825 nm, were the GNR (~6 nm long) got lost. A subsequent approach of the tip reveals no shift at all, as illustrated by the frequency shift in fig. S12.

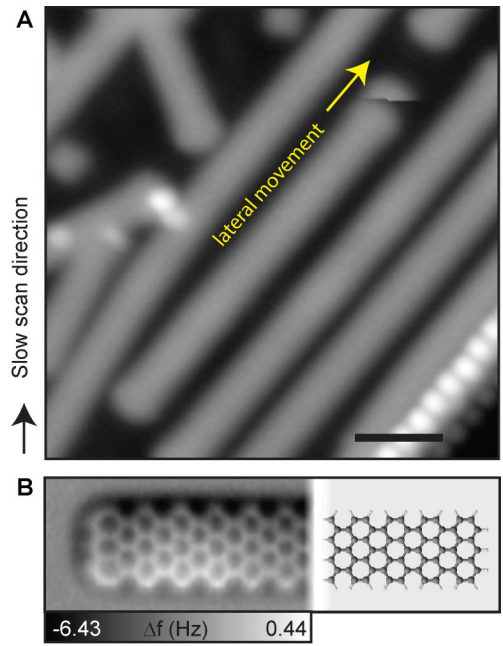

**Fig. S1.** GNR manipulation during imaging. (**A**) STM topography showing a tip-induced lateral manipulation of a N=7 GNR during scanning. Scale bar 1 nm, measurement parameters: tunneling current I=900 fA and bias voltage applied to the tip V =−200 mV (**B**) AFM image (left), compared to a schematic drawing of the chemical structure (right) of the GNR. In this case the tip oscillation amplitude A=43 pm.

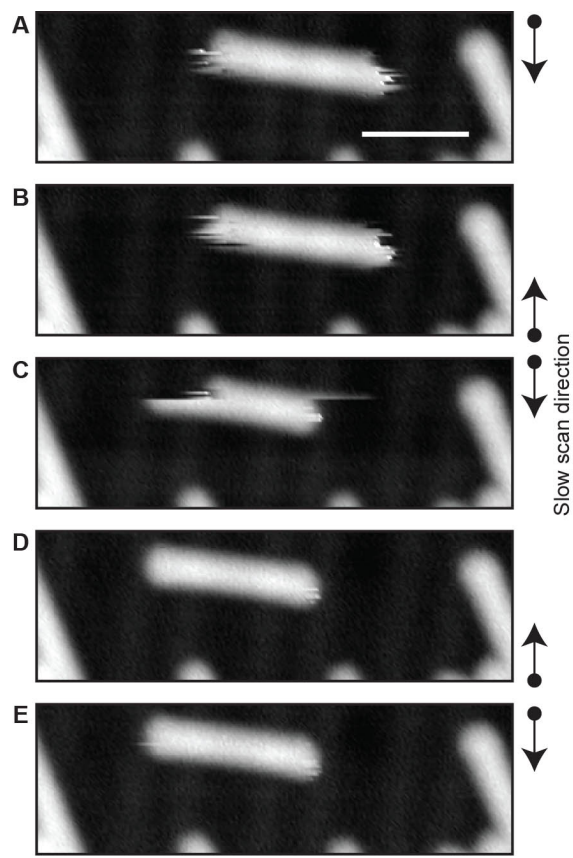

**Fig. S2.** Adsorption site-dependent diffusion barrier. **(A-E)** Consecutive scanning tunneling microscopy (STM) topographies of the same GNR. Scale bar 5 nm, measurement parameters $V$=-200 mV and $I$=1 pA.

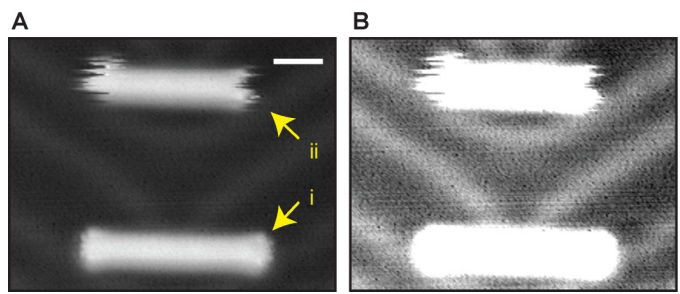

**Fig. S3.** Influence of the herringbone reconstruction on the diffusion barrier. (**A**) STM topography in which two parallel GNRs with slightly different length are found parallel and lying on the same region of the elbow reconstruction. (**B**) Same image with increased contrast to highlight the reconstruction. Scale bar 2 nm, measurement parameters $V$=-200 mV and $I$=2 pA.

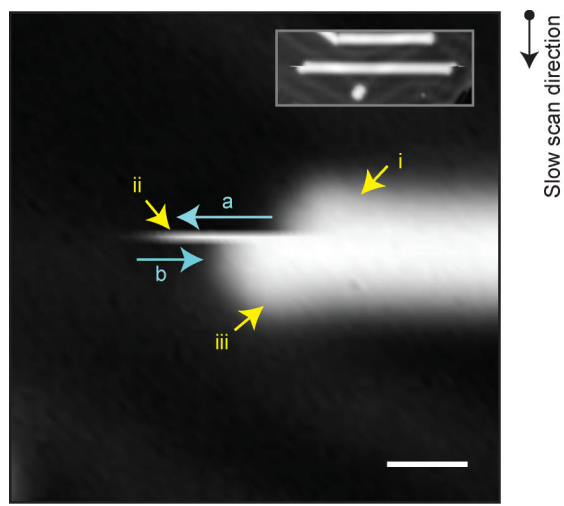

**Fig. S4.** Subsequent GNR slips in accidental tip-induced manipulation. STM image showing accidental manipulations along the [-1,0,1] direction. The GNR steady states are labelled (i), (ii) and (iii) respectively, and the slips by the blue arrows a and b. The inset shows the full GNR approximately 18.5 nm long (22 precursor molecules). Scale bar 1 nm, measurement parameters $V$=-200 mV and $I$=2 pA.

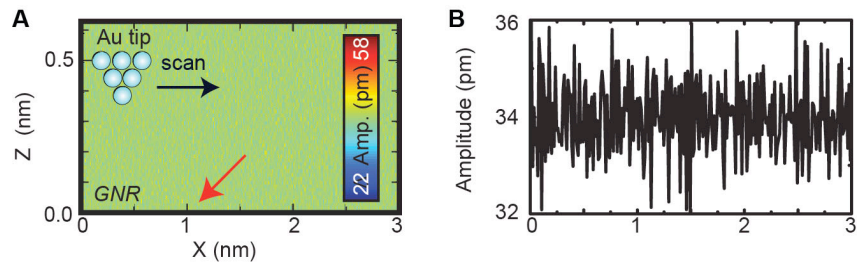

**Fig. S5.** Amplitude signals in static friction force measurement. (A) Two dimensional amplitude map along the longitudinal axis of the manipulated GNR. (B) Distance dependence of amplitude before, during and after the GNR displacement. The data was collected together with the Δf signal shown in Fig. 1B and 1C. Since the energy loss in the manipulation is less than the detection limit, no significant amplitude change was detected.

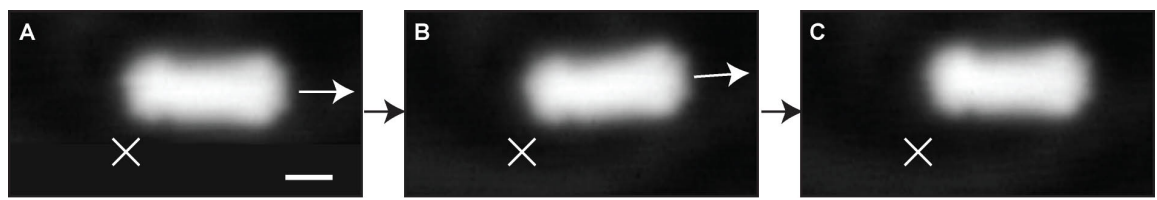

**Fig. S6.** Lateral manipulation of a short GNR. (**A**) STM image before lateral manipulation of a short GNR, composed of two precursor molecules in the direction indicated by the white arrow. (**B**) STM image taken after the first lateral manipulation. In contrast to the monomer of fig. S7, the GNR has been successfully manipulated laterally, but a rotation of approximately 4 degrees counter clockwise is also visible. In order to show how this rotation occurs systematically, a second manipulation along the GNR longitudinal axis has been performed. (**C**) STM image taken after the second lateral manipulation. The GNR results again displaced laterally and a second rotation of 4 degrees clockwise with respect to the manipulation direction brought it back to a perfectly horizontal position. Scale bar 1 nm, measurement parameters $V$=-200 mV and $I$=2 pA.

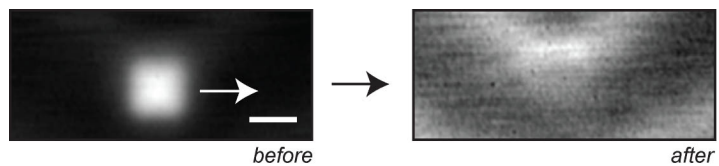

before → after

**Fig. S7.** Diffusion barrier versus desorption barrier. STM images before and after the manipulation of a short GNR formed by only one precursor molecule. Before the lateral manipulation started, the GNR was picked up by the tip, meaning that a vertical manipulation occurred. Scale bar 1 nm, measurement parameters $V$=-200 mV and $I$=2 pA.

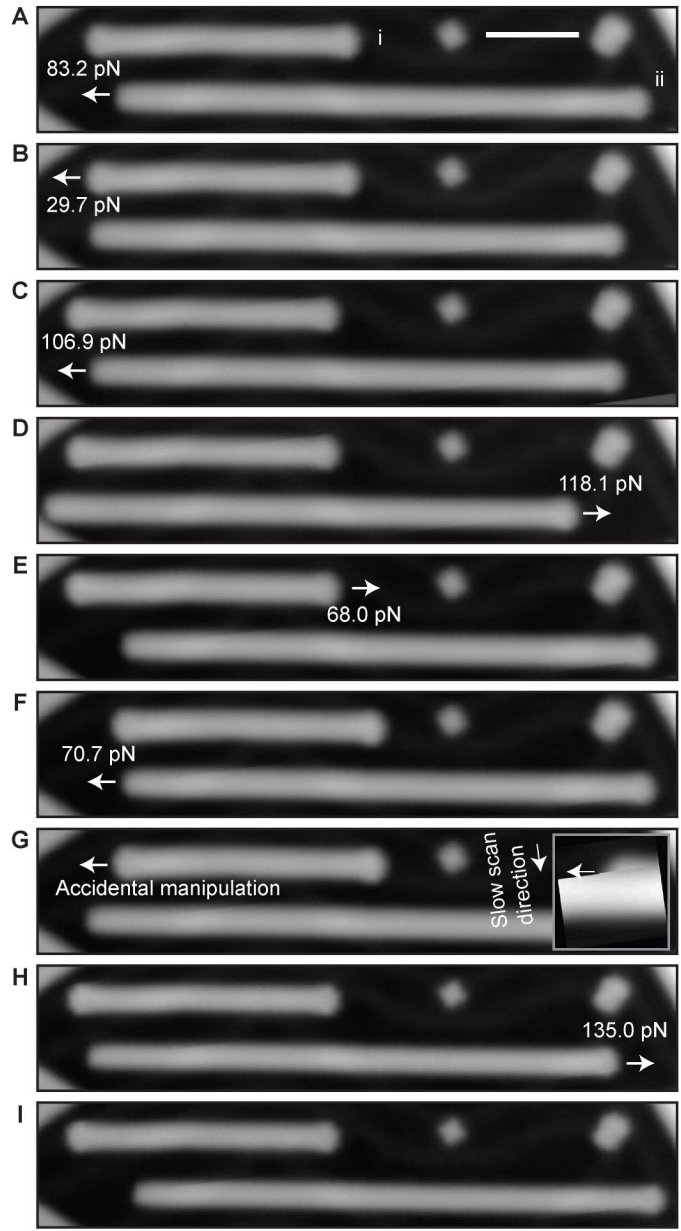

**Fig. S8.** A series of controlled tip-induced manipulations. (**A-I**) A series of STM taken after subsequent tip-induced manipulations. The lateral forces to move the GNRs are extracted via the measured two-dimensional frequency shift map as discussed in the main text. Scale bar 5 nm, measurement parameters $V$=-200 mV and $I$=2 pA.

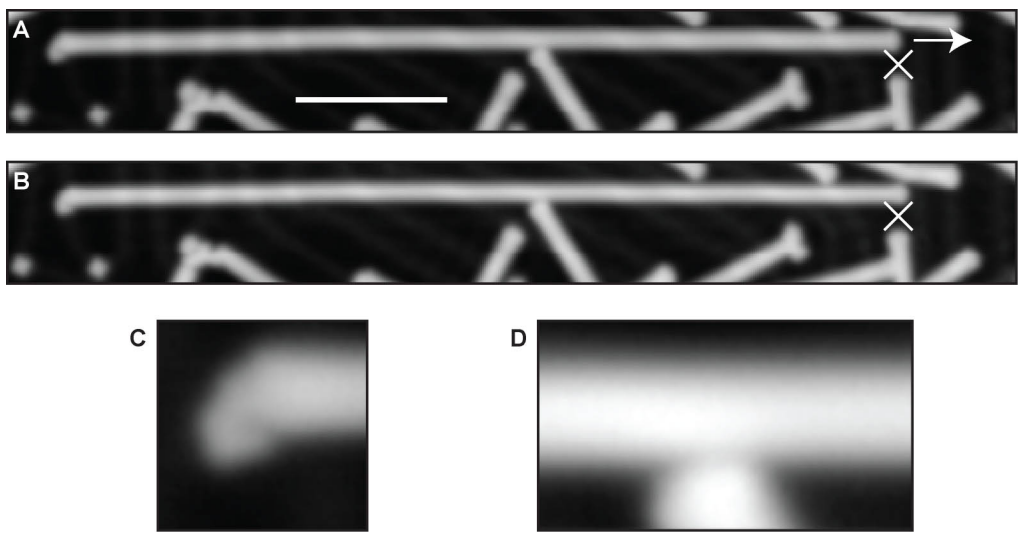

**Fig. S9.** Lateral manipulation of a long GNR. (**A**) STM image taken before lateral manipulation of a long GNR (55 nm) along the direction indicated by the arrow. (**B**) STM image taken after the lateral manipulation. The extracted lateral force is 211 pN, which is very large compared to shorter GNRs. (**C**) Magnified STM image around a terminus of the GNR. (**D**) Magnified STM image around the middle of the GNR. Scale bar 10 nm, measurement parameters $V$=-200 mV and $I$=2 pA.

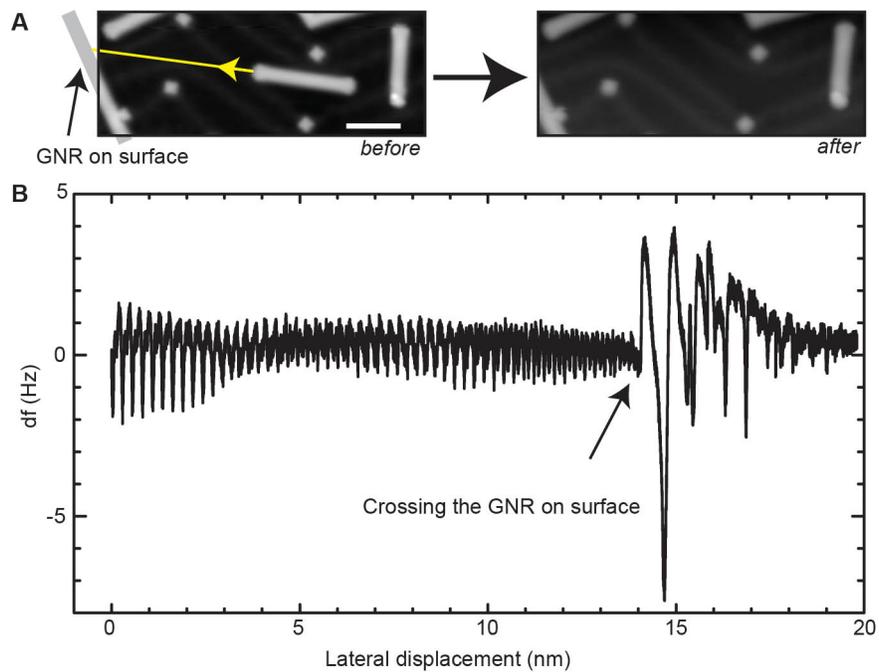

**Fig. S10.** Lateral manipulation of a GNR colliding against another GNR. (**A**) STM images before and after the lateral manipulation, scale bar 5 nm. (**B**) Frequency shift curve recorded during the manipulation.

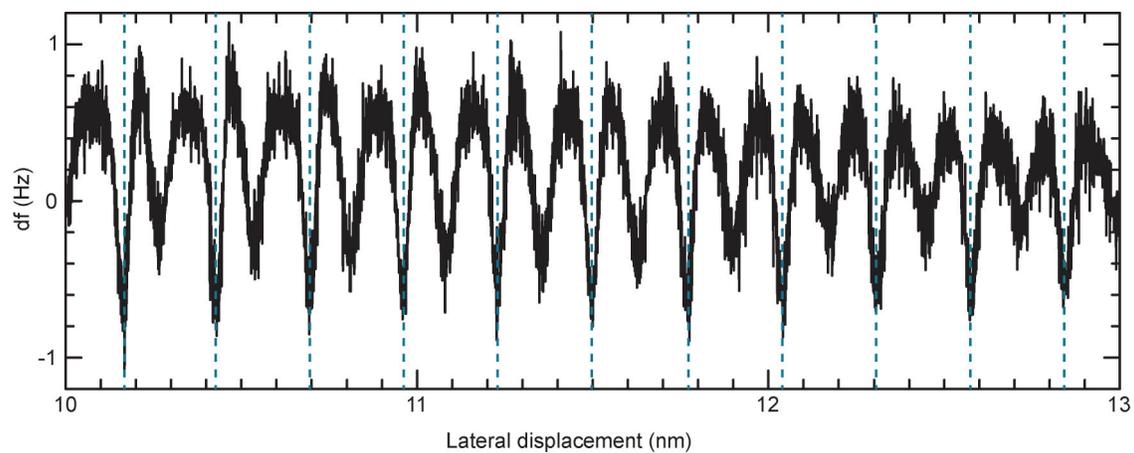

**Fig. S11.** Frequency shift at small tip-substrate separation $Z$. Magnification of the frequency shift reported in fig. S10, a wiggling of periodicity roughly half of the Au one is clearly visible, corresponding to 3→1→3 jumps.

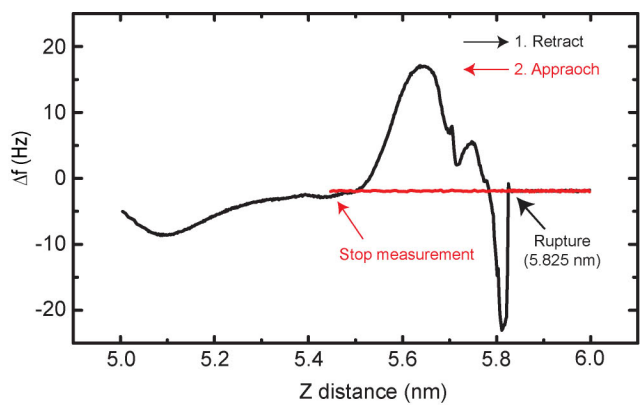

**Fig. S12.** Complete detachment of the GNR. Frequency shift recorded as a function of the tip-surface separations $Z$ during a retraction/approach procedure with $Z$ almost equal to the GNR length, here approximately 6 nm. For $Z=5.825$ nm an abrupt change of the frequency shift reveals the complete detachment of the GNR from the substrate. Notice that in such a condition the GNR is almost in vertical position with the short edge of the GNR contacting the surface located (in the *xy* plane) almost exactly below the tip apex (*23*). Since no significant frequency shift was detected in the subsequent approach we conclude the GNR is lost and, finding no evidence of its presence in a subsequent image of the surface, it remained presumably attached to the tip itself, in a region far away from the apex.

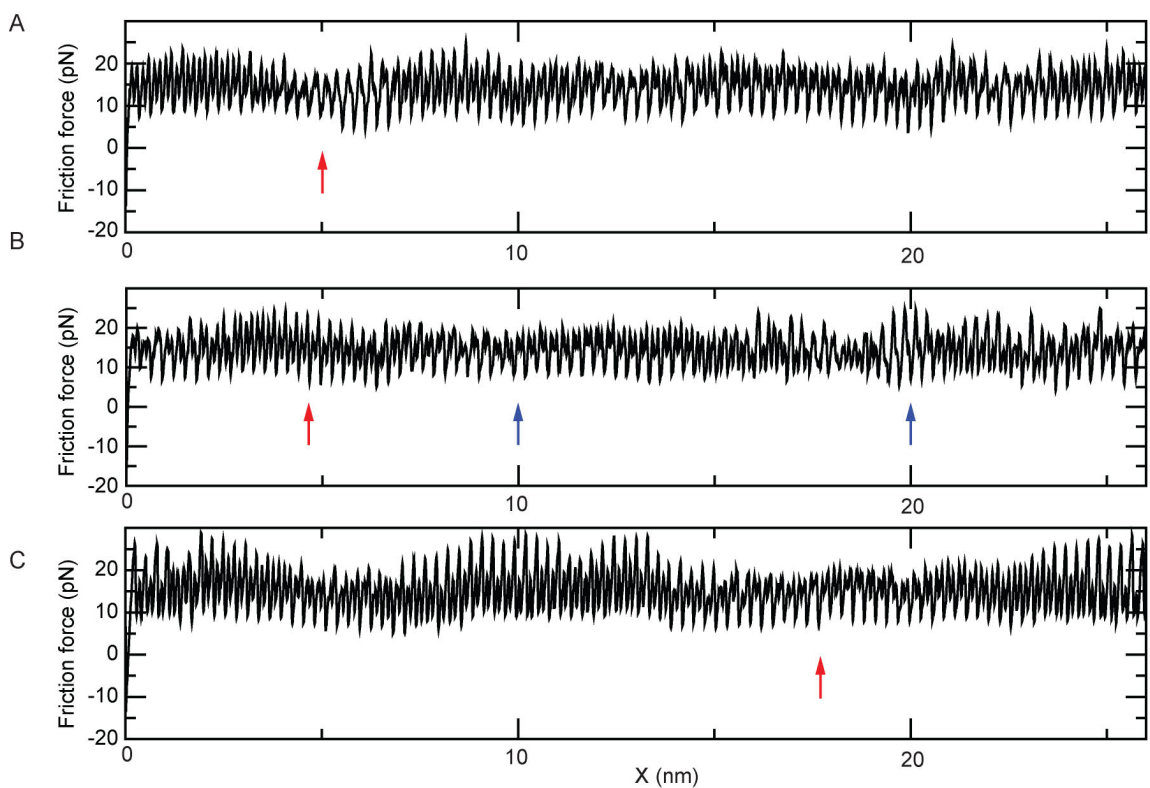

**Fig. S13.** Friction force and the Au reconstruction. The panels show the lateral (friction) force recorded during the simulation of GNR driven along the three scan lines presented in Fig. 3 of the main text. (**A-C**) correspond to the frequency shifts $\Delta f$ in Fig. 3E-G of the main text. A clear connection between the amplitude of $\Delta f$ oscillations and the intensity of the friction force is visible: the larger the $\Delta f$ oscillations the larger the friction force value (blue arrows). Regions where $\Delta f$ oscillates with small amplitude correspond to configurations of small friction force (red arrows).

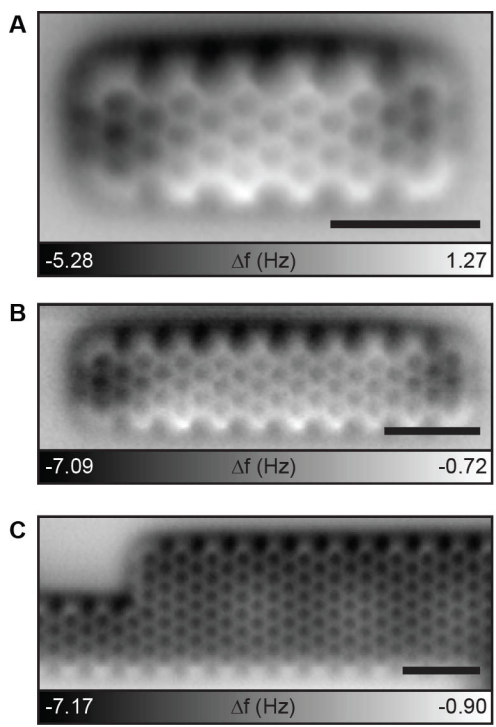

**Fig. S14.** GNRs with different lengths and widths on Ag(111). (**A-B**) AFM image of a N=7 GNR on Ag(111) obtained in constant height mode. Both termini were observed as dark (more negative frequency shift), meaning that the part adsorbed closer to the Ag(111) surface. (C) AFM image of a fused N=14 GNR. Moiré pattern was observed. These variations of the height indicated a stronger GNR-substrate interaction, compared to the case on Au(111). These GNRs on Ag(111) could not be manipulated with force values similar to those used on Au(111). Scale bar 1 nm, measurement parameters: Oscillation amplitude A=50 pm.

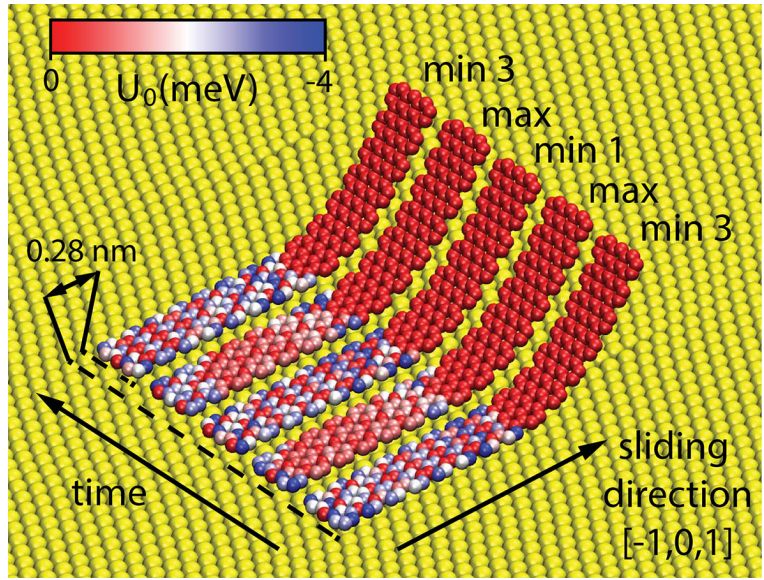

**Fig. S15.** Detailed dynamics of the lateral motion. Energy landscape experienced by the C atoms in the GNR during a double transition between the configurations 1 and 3 in Fig. 3C of the main text (configuration 2 being suppressed on a reconstructed substrate). "Min" and "max" refer to the time instants immediately following and preceding a slip.

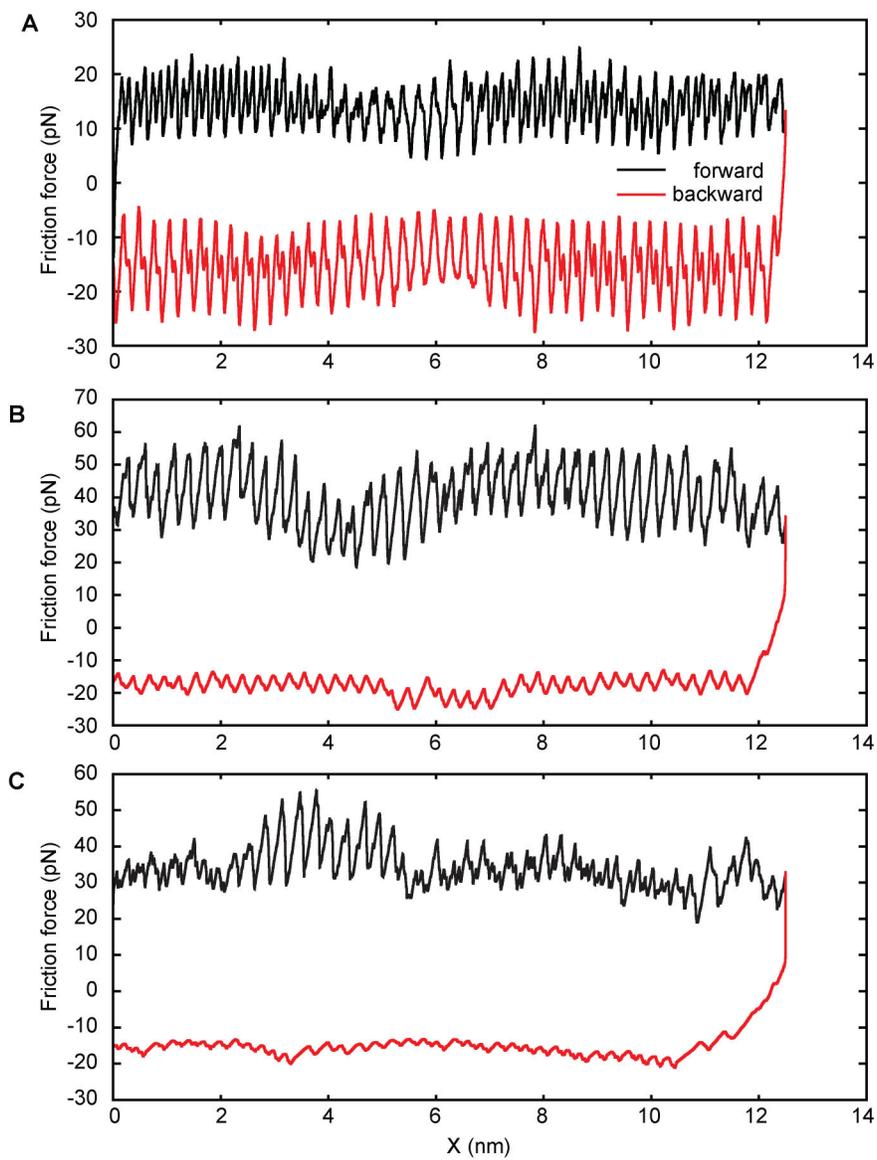

**Fig. S16.** Friction force loops corresponding to the simulated forward and backward scans in Fig. 4 of the main text for $Z$=2,3 and 4 nm.